\documentclass[doublecol]{epl2} 
\usepackage{graphicx}
\usepackage{amsmath}
\usepackage{amssymb}
\newcommand{\qmave}[1]{\langle #1 \rangle}
\newcommand{\ave}[1]{\overline{\qmave{#1}}}

\title{Bunching and anti-bunching of localised particles in disordered media}
\shorttitle{Bunching and anti-bunching of localised particles in disordered media} 

\author{Frank Schlawin \and Nicolas Cherroret \and Andreas Buchleitner}
\shortauthor{F. Schlawin \etal}

\institute{                    
 Physikalisches Institut, Albert-Ludwigs-Universit\"at Freiburg, Hermann Herder Stra\ss{}e 3, D-79104 Freiburg, Germany
}
\pacs{42.25.Dd}{Wave propagation in random media}
\pacs{42.50.-p}{Quantum Optics}
\pacs{72.15.Rn}{Localisation effects}

\abstract{
We consider pairs of non-interacting quantum particles transmitted through a disordered medium, with emphasis on the role of their quantum statistics. It is shown that particle-number correlations measured in transmission are strikingly sensitive to the quantum nature of the particles when they undergo Anderson localisation, due to bosonic bunching and fermionic anti-bunching in the scattering channels of the medium. The case of distinguishable particles is also discussed.
}

\begin{document}

\maketitle

Wave transport in disordered media has been a subject of intense research efforts for over 50 years. Interference phenomena such as Anderson localisation \cite{Anderson58}, coherent backscattering  \cite{vanAlbada85, Wolf85}, or universal conductance fluctuations \cite{Stone85, Skocpol86} represent striking examples of the impact of disorder on wave propagation. While these phenomena can be explained on the level of wave mechanics, the impact of the quantised nature of the propagating wave or of the scattering medium has recently moved into focus of theoretical and experimental investigations. For instance, for classical light scattered resonantly off ``quantum'' disordered media consisting of cold atoms, the atomic internal structure has dramatic effects on the visibility of the coherent backscattering effect \cite{Labeyrie99, Wellens10}. Conversely, the quantum nature of the propagating object can also come into play and lead to original phenomena, like in the context of cold atomic gases scattered off random optical potentials, where extensions of standard mean-field (wave-like) descriptions of matter waves allowed for instance to demonstrate Anderson localisation of the quantum excitations of Bose-Einstein condensates \cite{Lugan07a}. 

In optics,  quantum statistics of multiparticle interference effects move now into reach of multiple scattering experiments. Recent theoretical works notably include investigations of photon noise and  photon correlations in conservative \cite{Lodahl05, Smolka09}, absorbing \cite{Beenakker00, Patra00} or amplifying \cite{Patra00, Fedorov09} disordered media, or even of quantum entanglement in the presence of disorder \cite{Beenakker09, Ott10, Cherroret11}. Most of these previous works considered transport of non-classical light in \emph{diffusive} media, i.e. where interferences are smeared out by a disorder average, but recently also the scattering of pairs of non-interacting quantum particles in one-dimensional disordered lattices was investigated under conditions of Anderson localisation \cite{Lahini10}. In particular, in this work the role of the quantum statistics of individual particles was pointed out, and bosonic bunching as well as fermionic anti-bunching effects were observed in particle-number correlations (known as Hanbury Brown-Twiss correlations in the absence of disorder \cite{Brown56}).

In the present Letter we present an original transport study of pairs of non-interacting quantum particles in waveguide-shaped media with \emph{three-dimensional} disorder, where a cross-over from diffusion to Anderson localisation is present. Our purpose is here to clarify how the quantum statistics of the particles (bosons, fermions or distinguishable particles) manifests itself in the particle-number correlation \emph{as the regime of transport is changed} inside the medium. From both a numerical and an analytical study based on the random matrix theory of quantum transport, we find that, as long as propagation is diffusive in the disordered medium, the correlation is hardly sensitive to the quantum statistics of the particles. In strong contrast, when Anderson localisation comes into play, the correlation exhibits completely different behaviour for bosons, fermions and distinguishable particles, due to bunching and anti-bunching effects in the scattering channels supported by the medium. 

\section{Model}

\begin{figure}
{\includegraphics[scale=1.2]{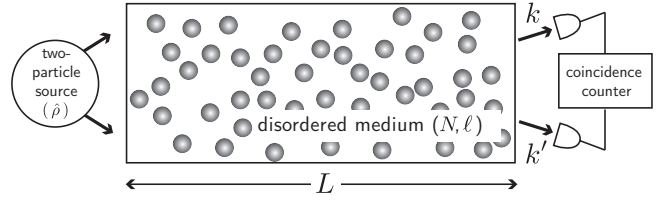}}
\caption{A source produces pairs of particles characterised by a two-particle density matrix $\hat{\rho}$. These pairs are scattered from a disordered medium with length $L$, mean free path $\ell$, and supporting $N$ transverse modes. The coincidence rate between two outgoing modes $k$ and $k^\prime$ is analysed in transmission.
\label{fig:setup}}
\end{figure}

We consider a source producing pairs of monochromatic particles with identical wave number $k_0$. These pairs propagate through a medium containing heterogeneities randomly arranged in space and characterised by a scattering mean free path $\ell$, see fig. \ref{fig:setup}. The medium has length $L$ and cross-section $A$, and is assumed to be connected to perfect leads (not shown in fig. \ref{fig:setup}). In such a geometry the incoming and outgoing wave fronts can be decomposed on a discrete set of $N\simeq k_0^2A$ transverse modes, each of them corresponding to a value of the transverse component of the wave vector compatible with the boundary conditions. The 
statistical properties of transport observables are independent of the transverse shape of the medium in the limit $N\gg1$ that we assume from here on. Particles' coincidences between two modes $k\ne k^\prime$ are detected in transmission. Such a single-mode detection setup requires a detector area smaller than the typical size $R^2/N$ of a speckle spot, with $R$ the distance between sample and detector. 
To quantify the coincident count statistics, we quantise the field before and after the disordered region, using the basis of the transverse modes \cite{Patra00}, and define the normalised correlation function
\begin{eqnarray}
C = \frac{\ave{:\! \hat{n}_k \hat{n}_{k'}\!\!:}}{\ave{\hat{n}_k} \times \ave{\hat{n}_{k'}}} \label{eq:correlation},
\end{eqnarray}
where $\hat{n}_{k}$ denotes the particle-number operator in the outgoing mode $k$, $\qmave{\ldots}$ the quantum expectation value, $:\,\,:$ the normal ordering of operators, and the overbar the average over a statistical ensemble of disordered samples. Since we are here primarily interested in the quantum statistics of the particles (fermions, bosons or distinguishable particles), we remain general in the following and do not specify how the two-particle state is initially distributed over the incoming modes. In the spirit of \cite{Beenakker09}, we thus write the initial density matrix in the general form 
\begin{eqnarray}
\hat{\varrho} = \frac{1}{2}\!\sum_{i,j,i^\prime,j^\prime} \! w_{i j i^\prime j^\prime}\, \hat{a}^{\dagger}_{i} \hat{a}^{\dagger}_{j} \vert 0 \rangle \langle 0 \vert \hat{a}_{i^\prime} \hat{a}_{j^\prime},
\label{eq:stateindist}
\end{eqnarray}
where the operator $\hat{a}^{\dagger}_{i}$ creates a particle in the transverse mode labeled by index $i$, and all sums run from $1$ to $N$. Creation and annihilation operators fulfill the commutation relations $\hat{a}^{\dagger}_{i}\hat{a}^{\dagger}_{j}=\pm\hat{a}_{j}^{\dagger}\hat{a}_i^{\dagger}$, $\hat{a}_{i}\hat{a}_{j}=\pm\hat{a}_{j}\hat{a}_i$ and $\hat{a}^{\dagger}_{i}\hat{a}_{j}=\delta_{ij}\pm\hat{a}_{j}^{\dagger}\hat{a}_i$, with $+$ for bosons and $-$ for fermions. Due to these relations, the tensor $w_{iji^\prime j^\prime}$ can be chosen (anti-) symmetric under the exchange of the first and last two indices for (fermions) bosons. The prefactor $1/2$ in eq. (\ref{eq:stateindist}) guarantees the state normalization, $\text{tr}\hat{\rho}=\sum_{i,j}w_{ijij}=1$. 
From the initial density matrix $\hat{\rho}$, the mean particle number in some outgoing mode $k$ is defined as
\begin{equation}
\qmave{\hat{n}_k}=\text{tr}[\hat{\rho}\,\hat{c}_k^\dagger(\underline{a}^\dagger)\hat{c}_k(\underline{a})],
\label{eq:state}
\end{equation}
where $\hat{c}_k(\underline{a})=\sum_i t_{ki}\hat{a}_i$ and $\hat{c}^\dagger_k(\underline{a}^\dagger)=\sum_i t_{ki}^*\hat{a}_i^\dagger$ are the input-output relations connecting the annihilation (creation) operator $\hat{c}_k$ ($\hat{c}_k^\dagger$) of the outgoing mode $k$ to the set of operators $\hat{a}_i$ ($\hat{a}_i^\dagger$) of the incoming modes $i$ through the (random) transmission matrix $t$ of the disordered medium. Substituting these relations into eq. (\ref{eq:state}), we readily obtain
\begin{equation}
\qmave{\hat{n}_k}=2\sum_{i,j,i^\prime}w_{iji^\prime j}t_{ki}t^*_{ki^\prime}.
\label{eq:nk}
\end{equation}
Similarly, from the definition 
\begin{equation}
\qmave{:\!\hat{n}_k\hat{n}_{k^\prime}\!\!:}=\text{tr}[\hat{\rho}\,\hat{c}_k^\dagger(\underline{a}^\dagger)\hat{c}_{k^\prime}^\dagger(\underline{a}^\dagger)\hat{c}_k(\underline{a})\hat{c}_{k^\prime}(\underline{a})]
\label{state}
\end{equation}
of the particle-number correlation, we obtain 
\begin{equation}
\qmave{:\!\hat{n}_k\hat{n}_{k^\prime}\!\!:}=2\!\sum_{i,j,i^\prime,j^\prime}w_{iji^\prime j^\prime}t_{ki}t^*_{k^\prime j}t^*_{ki^\prime}t_{k^\prime j^\prime}.
\label{eq:nknkp}
\end{equation}
Note that at this stage eqs. (\ref{eq:nk}) and (\ref{eq:nknkp}) hold for fermions \emph{and} bosons, provided the tensor $w_{iji^\prime j^\prime}$ is properly (anti-) symmetrised. 

The next step in the derivation of the correlation function (\ref{eq:correlation}) consists in performing the ensemble averages. In eqs. (\ref{eq:nk}) and (\ref{eq:nknkp}), statistical properties of the disordered medium are encoded in the transmission matrix elements $t_{ki}$. In the limit $N\ell/L\equiv g\gg1$, which corresponds to a regime where particles are multiply scattered according to a diffusion process, these elements are normally distributed over the unitary group \cite{Beenakker97}. This property was notably used in \cite{Beenakker09} to calculate the statistical distribution of $\qmave{:\!\hat{n}_k\hat{n}_{k^\prime}\!\!:}$ for photons. 

Here however, we aim at studying the behaviour of $C$ for \emph{arbitrary} values of the parameter $g$. For this purpose we make use of the unitarity and symmetry of the transmission matrix, which allows us to employ the polar decomposition \cite{Beenakker97, Mello88}
\begin{eqnarray}
t_{k i} = \sum_{a } u_{k a} \sqrt{\tau_a} v_{a i},
\label{eq:polard}
\end{eqnarray}
where $u$ and $v$ are random unitary matrices and $\tau_1, \dots, \tau_N$ are the eigenvalues of $t$. 
Eq. (\ref{eq:polard}) is at the basis of the macroscopic description of disordered media \cite{Beenakker97}. Physically, the matrix $v$ first distributes a field amplitude incident in mode $i$ among the \emph{scattering channels} of the medium, each having its transmission coefficient $\tau_a$ (between 0 and 1). The matrix $u$ finally recombines the transmitted field amplitudes in the outgoing mode $k$. 

For a medium with length much larger than its width, i.e in the limit $L\gg\sqrt{A}$, ensemble averaging can be carried out in two steps: first over the matrices $u$ and $v$, which are uniformly distributed over the unitary group (isotropy assumption), and second over the eigenvalues $\tau_1, \dots, \tau_N$, whose joint distribution follows the so-called DMPK equation \cite{Mello88}. Substituting the decomposition (\ref{eq:polard}) into eqs. (\ref{eq:nk}) and (\ref{eq:nknkp}), and performing the average over unitary matrices, we obtain
\begin{eqnarray}
C = \frac{1}{2} \frac{\sum_{a,b}\overline{ \tau_a\tau_b}\pm\sum_a\overline{\tau_a^2}}{\left(\sum_a\overline{\tau_a}\right)^2}, 
\label{eq:correlationfb}
\end{eqnarray}
where the positive sign refers to bosons and the negative sign to fermions. Eq. (\ref{eq:correlationfb}) is the first important result of this Letter. It shows that, when $L\gg\sqrt{A}$, the correlation function $C$ is independent of how the incident state is distributed over the incident modes of the disordered medium, and only depends on the statistical correlations between scattering channels \cite{Beenakker09}. This property originates from the fact that when deriving eq. (\ref{eq:correlationfb}), we assumed each field amplitude  
incoming in a given mode to be equally distributed among the scattering channels of the medium (isotropy assumption). Another important comment concerns the prefactor $1/2$ in eq. (\ref{eq:correlationfb}), also found in \cite{Ott10, Cherroret11}. For the particular case of a two-photon Fock state with Fano factor $F=0$, this prefactor was shown to express itself as $1+(F-1)/2$ \cite{Smolka09}. Its value $1/2<1$ therefore signals the sub-Poissonian statistics, here of an arbitrary incident two-particle state.

We now comment on the structure of the mean correlation [numerator of eq. (\ref{eq:correlationfb})]. The latter consists of two terms. The first one, $\sum_{a,b}\overline{ \tau_a\tau_b}$, corresponds to the sum of all possible correlations between scattering channels. The second one, $\pm\sum_{a}\overline{\tau_a^2}$, corresponds to the sum of all possible autocorrelations of scattering channels and is affected by the specific statistics of quantum particles: fermions cannot both propagate in the same channel, therefore this sum appears with a minus sign, and cancels the autocorrelation contributions in the first term (anti-bunching effect). On the other hand, bosons tend to bunch in the scattering channels, as manifested by the positive sign in front of the second term. These bunching and anti-bunching effects 
have important consequences on the magnitude of the correlation function $C$, as we now show.

\section{Numerics for bosons and fermions}

Eq. (\ref{eq:correlationfb}) is valid for any value of the parameter $g=N\ell/L$, provided $L\gg\sqrt{A}$. In order to clarify how bunching and anti-bunching effects would show up concretely in an experiment accessing the correlation function (\ref{eq:correlation}), we evaluated $C$ numerically as a function of $L/\ell$, for a fixed number of modes, $N =10$. For this purpose we computed each of the averages $\sum_a\overline{\tau_a^2}$, $\sum_{a,b}\overline{ \tau_a\tau_b}$ and $\sum_a\overline{\tau_a}$ from the DMPK equation, by means of a Markov chain Monte Carlo approach \cite{Krauth06}. Such a numerical procedure was already used in the context of the conductance distribution of metallic conductors \cite{Froufe02} and of intensity correlations of classical light in disordered media \cite{Cwilich06}. 
\begin{figure}[h]
{\includegraphics[scale=0.86]{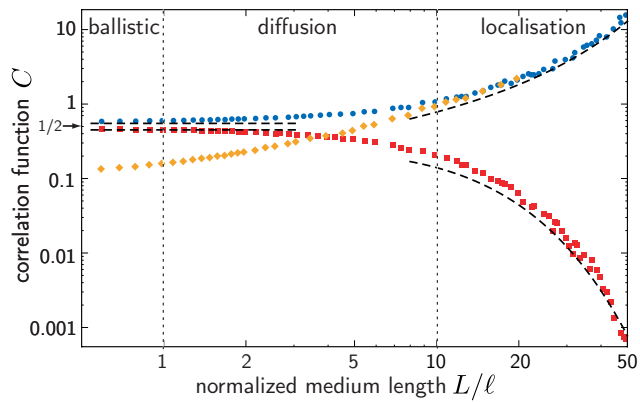}}
\caption{Correlation function (\ref{eq:correlationfb}) for bosons (dots, blue online) and fermions (squares, red online), plotted as functions of $L/\ell$, for $N=10$. Diamonds (orange online) are the contribution $\sum_a\overline{\tau_a^2}/\sum_a\overline{\tau_a}^2$ to $C^\text{bosons}$. Dashed horizontal lines are the analytical prediction (\ref{eq:ballistic}), and dashed curves are given by eqs. (\ref{eq:locb}) and (\ref{eq:locf}). Vertical dotted lines indicate the crossovers $L\sim\ell$ and $L\sim N\ell$ from ballistic to diffusion and from diffusion to localisation, respectively.
\label{fig:ind_part}}
\end{figure}
At fixed $N$, the study of $C$ as a function of $L/\ell$ allows us to probe all possible regimes of transport in the disorder, from quasi-ballistic ($L\lesssim\ell$) to diffusive propagation ($\ell\ll L\ll N\ell$) and eventually Anderson localisation ($L\gg N\ell$). The results can be seen in fig. \ref{fig:ind_part}, for bosons (dots, blue online) and fermions (squares, red online). In the quasi-ballistic and diffusive regimes, $C$ is very close to $1/2$ for bosons and fermions, i.e. the correlation is hardly sensitive to the quantum statistics of the particles (the small difference observed in fig. \ref{fig:ind_part} is discussed in the next section). This picture dramatically changes as $L$ increases, at the onset of localisation ($L\sim N\ell$): the bosonic correlation function starts growing, while the fermionic counterpart decreases. Far in the localisation regime ($L\gg N\ell$), $C\gg1$ for bosons and $C\ll1$ for fermions. These radically different behaviours can be interpreted in the framework of the theory of active transmission channels \cite{Imry86, Stone91} by rewriting eq. (\ref{eq:correlationfb}) as $C^\text{bosons}= (1/2)\times\left(\sum_{a\ne b}\overline{ \tau_a\tau_b}+2\sum_a\overline{\tau_a^2}\right)/\left(\sum_a\overline{\tau_a}\right)^2$ and $C^\text{fermions}= (1/2)\times\left(\sum_{a\ne b}\overline{ \tau_a\tau_b}\right)/\left(\sum_a\overline{\tau_a}\right)^2$. As the length $L$ increases, more and more scattering channels become closed, $i.e.$ their transmission coefficient becomes exponentially small. Eventually, in the localisation regime, only one channel $a_0$ retains a fairly large transmission coefficient. In this limit, we thus have $\left(\sum_a\overline{\tau_a}\right)^2\simeq \overline{\tau_{a_0}}^2$, $\sum_a\overline{\tau_a^2}\simeq \overline{\tau_{a_0}^2}$ and $\sum_{a\ne b}\overline{ \tau_a\tau_b}\simeq \overline{ \tau_{a_0}\tau_{b_0}}$, where $b_0$ denotes a closed channel. Since the correlation between a closed and the open channel is typically very small compared to $\overline{\tau_{a_0}}^2$, we have  $C^\text{fermions}\ll 1$. On the other hand, we have also $\overline{\tau_{a_0}\tau_{b_0}}\ll \overline{\tau_{a_0}^2}$, such that $C^\text{bosons}\simeq(1/2)\times\overline{\tau_{a_0}^2}/\overline{\tau_{a_0}}^2$. This ratio is very large because transmission fluctuations well exceed the mean transmission in the localisation regime \cite{Beenakker97}. To get a better picture of the weight of each term appearing in $C^\text{bosons}$ and $C^\text{fermions}$, in the different regimes of transport, we also show in fig. \ref{fig:ind_part} the contribution of channel autocorrelations to $C^\text{bosons}$, $\sum_a\overline{\tau_a^2}/\sum_a\overline{\tau_a}^2$ (diamonds, orange online).

It is instructive to compare the results in fig. \ref{fig:ind_part} for bosons (dots) with those obtained by Ott \emph{et al.}, fig. 4 of Ref. \cite{Ott10}, for the particular case of a two-photon Fock state. Our correlation function is insensitive to the form of the incident state, unlike in \cite{Ott10}, where the correlation function is constant for two photons incident in the same mode whereas it increases and saturates for two photons incident in two different modes. This difference relies on the different quantity appearing in the denominator of the correlation function of \cite{Ott10}, $\overline{\qmave{n_k}\qmave{n_k^\prime}}$, instead of $\overline{\qmave{n_k}}\times\overline{\qmave{n_k^\prime}}$ in our case, eq. (\ref{eq:correlation}).

\section{Analytical results}

In order to support our numerical observations, we now calculate explicitly $C$ in the quasi-ballistic, diffusive and localisation regimes. First, when $L\lesssim\ell$, we have $\tau_a\simeq1\, \forall a$, such that
\begin{equation}
C\simeq\frac{1}{2} \frac{N^2\pm N}{N^2}=\frac{1}{2}\left(1\pm\frac{1}{N}\right). 
\label{eq:ballistic}
\end{equation} 
Eq. (\ref{eq:ballistic}) is shown in fig. \ref{fig:ind_part} for bosons ($+$) and fermions ($-$) as dashed horizontal lines. The $1/N$ difference between the two types of particles is visible in the figure since we considered a finite value of $N$ for the numerics. In practice however, $N\gg1$ and this difference is negligible. 

In the diffusive regime $\ell\ll L\ll N\ell$, averages over transmission eigenvalues can be evaluated by making use of the method of moments introduced in \cite{Mello91}. The result is a perturbation expansion of $C$ in powers of the parameter $1/g=L/(N\ell)\ll1$:
\begin{eqnarray}
C= \frac{1}{2} \left[1\pm\frac{2}{3g}+\mathcal{O} \left(\frac{1}{g^2}\right)\right].
\label{eq:diffusive}
\end{eqnarray} 
In the regime of diffusive transport, bosonic bunching ($+$) and fermionic anti-bunching ($-$) in the scattering channels are thus visible, but remain small as long as $g\gg1$. 

Finally, far in the localisation regime $L\gg N\ell$, $C$ can be calculated from the probability distribution of the transmission eigenvalues \cite{Beenakker97}. A straightforward calculation then yields:
\begin{eqnarray}
C^\text{bosons} = \frac{1}{3} \sqrt{\frac{\pi L}{2\xi}} \exp \left( \frac{L}{2 \xi} \right), \label{eq:locb}\\
C^\text{fermions} = \frac{1}{2} \sqrt{\frac{\pi L}{2 \xi}} \exp \left( - \frac{3 L}{2\xi} \right), \label{eq:locf}
\end{eqnarray}
where we have explicitly introduced the localisation length $\xi=N\ell$. Eqs. (\ref{eq:locb}) and (\ref{eq:locf}) are consistent with the discussion above, namely $C^\text{fermions}\ll1\ll C^\text{bosons}$. Analytical predictions (\ref{eq:locb}) and (\ref{eq:locf}) are shown in fig. \ref{fig:ind_part} as dashed curves, and agree  well with our Monte Carlo simulations.

\section{Distinguishable particles}

To complete our discussion, we finally consider the scattering problem of a pair of \emph{distinguishable} particles. For this purpose, we equip them with an additional spin degree of freedom, indicated by the symbols $\uparrow$ and $\downarrow$. With this strategy, we consider an incident state of the form 
\begin{eqnarray}
\hat{\rho} =\sum_{i,j,i^\prime,j^\prime} \! w_{i j i^\prime j^\prime}\, \hat{a}^{\dagger}_{i\uparrow} \hat{a}^{\dagger}_{j\downarrow} \vert 0 \rangle \langle 0 \vert \hat{a}_{i^\prime\uparrow} \hat{a}_{j^\prime\downarrow}.
\label{eq:statedist}
\end{eqnarray}
In comparison with eq. (\ref{eq:stateindist}), note the missing prefactor $1/2$. This originates from the new commutation relation involving the two particles which have now different spins, $\hat{a}_{k\uparrow}\hat{a}_{k\downarrow}^\dagger=\hat{a}_{k\downarrow}^\dagger\hat{a}_{k\uparrow}$. Normalization of  the state ($\ref{eq:statedist}$) again reads $\text{tr}\hat{\rho}=\sum_{i,j}w_{ijij}=1$. 

\begin{figure}
{\includegraphics[scale=0.85]{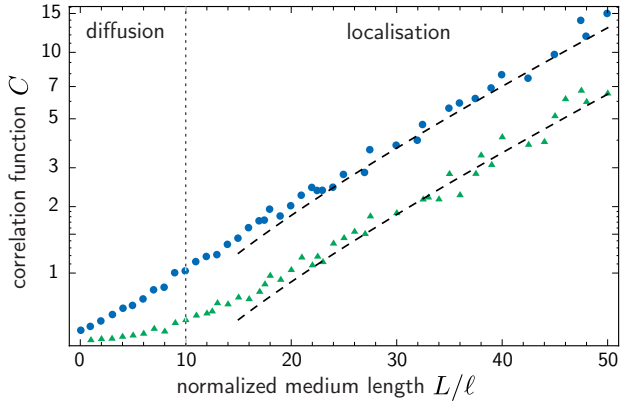}}
\caption{Correlation function for bosons (dots, blue online) and distinguishable particles (triangles, green online), plotted as functions of $L/\ell$ for $N=10$. The upper and lower dashed curves are the analytical predictions (\ref{eq:locb}) and (\ref{eq:correlationd}) respectively, and the vertical dotted line indicates the crossover $L\sim N\ell$ from diffusion to localisation.
\label{fig:Cdist}}
\end{figure}

The formalism developed above for indistinguishable particles remains essentially the same for distinguishable particles, but care has to be taken in defining the particle-number and correlation operators. Indeed, since one detects particles on output of the medium irrespective of their spin, these operators must now be symmetrised with respect to the two particles, i.e
\begin{equation}
\hat{n}_k=\hat{c}_{k\uparrow}^\dagger\hat{c}_{k\uparrow}+\hat{c}_{k\downarrow}^\dagger\hat{c}_{k\downarrow}
\end{equation}
for the correlation operator in the outgoing mode $k$, and
\begin{equation}
:\!\hat{n}_k\hat{n}_{k^\prime}\!\!:=\hat{c}_{k\uparrow}^\dagger\hat{c}_{k^\prime\downarrow}^\dagger\hat{c}_{k\uparrow}\hat{c}_{k^\prime\downarrow}+
\hat{c}_{k\downarrow}^\dagger\hat{c}_{k^\prime\uparrow}^\dagger\hat{c}_{k\downarrow}\hat{c}_{k^\prime\uparrow}.
\end{equation}
for the coincidence rate operator between the outgoing modes $k$ and $k^\prime$. The input-output relations now read $\hat{c}_{k\uparrow(\downarrow)}=\sum_i t_{ki}^{\uparrow(\downarrow)}\hat{a}_{i\uparrow(\downarrow)}$ and $\hat{c}^\dagger_{k\uparrow(\downarrow)}=\sum_i t_{ki}^{*\uparrow(\downarrow)}\hat{a}_{i\uparrow(\downarrow)}^\dagger$ (note the spin labeling also in the transmission matrix elements). Quantum expectation values for $\hat{n}_k$ and $:\!\hat{n}_k\hat{n}_{k^\prime}\!\!:$ are now given by 
\begin{equation}
\qmave{\hat{n}_k}=\sum_{iji^\prime}w_{iji^\prime j} t_{ki}^{\uparrow}t_{ki^\prime}^{*\uparrow}+
\sum_{ijj^\prime}w_{ijij^\prime} t_{kj}^{\downarrow}t_{kj^\prime}^{*\downarrow},
\label{eq:nk_d}
\end{equation}
and
\begin{equation}
\qmave{:\!\hat{n}_k\hat{n}_{k^\prime}\!\!:}=\sum_{iji^\prime i^\prime}w_{iji^\prime j^\prime} 
\left(t_{ki}^{\uparrow}t_{ki^\prime}^{*\uparrow}
t_{k^\prime j}^{\downarrow}t_{k^\prime j^\prime}^{*\downarrow}+
t_{k^\prime i}^{\uparrow}t_{k^\prime i^\prime}^{*\uparrow}
t_{k j}^{\downarrow}t_{k j^\prime}^{*\downarrow}
\right).
\label{eq:nknkp_d}
\end{equation}
Note that in eq. (\ref{eq:nk_d}), the first (second) term on the right-hand side is nothing but the contribution of the particle with spin up (down) to the detected mean particle number in mode $k$. Similarly, the two terms on the right-hand side of eq. (\ref{eq:nknkp_d}) correspond to the two possible coincidence events, i.e. particle with spin up detected in mode $k$ and particle with spin down detected in mode $k^\prime$ for the first term, and vice versa for the second one. Finally, we again decompose transmission coefficients according to $t_{k i}^{\uparrow(\downarrow)} = \sum_{a = 1}^N u_{k a}^{\uparrow(\downarrow)} \sqrt{\tau_a} v_{a i}^{\uparrow(\downarrow)}$ and carry out the averages over the matrices $u^{\uparrow,\downarrow}$ and $v^{\uparrow,\downarrow}$. This yields:
\begin{eqnarray}
C^\text{disting} = \frac{1}{2} \frac{\sum_{a,b}\overline{ \tau_a\tau_b}}{\left(\sum_a\overline{\tau_a}\right)^2}.
\label{eq:correlationd}
\end{eqnarray}
Comparing with eq. (\ref{eq:correlationfb}), we see that the bunching/anti-bunching term has disappeared for distinguishable particles, as expected. We computed  $C^\text{disting}$ numerically, using the same Monte Carlo approach as above to carry out the remaining averages  over transmission eigenvalues. The results are shown in fig. \ref{fig:Cdist} as a function of $L/\ell$ (triangles, green online). For comparison we also show $C^\text{bosons}$ (dots, blue online).
As for bosons, the correlation function of distinguishable particles increases with $L/\ell$. This is expected since at the onset of localisation, one scattering channel $a_0$ dominates transport and $\sum_{a,b}\overline{\tau_a\tau_b}\simeq\overline{\tau_{a_0}^2}\gg(\sum_{a}\overline{\tau_a})^2\simeq\overline{\tau_{a_0}}^2$. Again, simple analytical results can be obtained in the quasi-ballistic regime of transport, where $C^\text{disting}=1/2$, and in the diffusion regime, where $C^\text{disting}=(1/2)\times\left[1+2/(15g^2)+\mathcal{O}(1/g^3)\right]$. On the other hand, far in the localisation regime we find
\begin{equation}
C^\text{disting}=\dfrac{1}{6}\sqrt{\dfrac{\pi L}{2\xi}}\exp\left(\dfrac{L}{2\xi}\right).
\end{equation}
This result is shown in fig. \ref{fig:Cdist} (lower dashed curve), and agrees well with the numerical calculations. It also confirms that the bunching effect of bosons is only fully visible in the localisation regime where
\begin{equation}
\frac{C^\text{bosons}}{C^\text{disting}} \stackrel{L \gg N\ell}{\longrightarrow} \; 2,
\label{eq:bunching}
\end{equation}
as also demonstrated by the numerical points in fig. \ref{fig:Cdist}.

\section{Conclusion} We have shown that particle-number correlations measured on output of a disordered medium are extremely sensitive to the quantum statistics of localised particles. This phenomenon reflects the bosonic bunching and fermionic anti-bunching in the scattering channels of the medium. An interesting extension of this work would be to analyse the scattering problem of many-particle quantum states.

\acknowledgments
We thank M. C. Tichy for useful discussions and S. E. Skipetrov for his comments on the manuscript. N. C. acknowledges financial support from the Alexander von Humboldt Foundation.

\end{document}